# Unravelling the coupling between excitonic quasiparticles-electron-phonon and role of interlayer coupling in vertically and horizontally aligned layered MoS$_2$

Deepu Kumar[1#], Rahul Kumar[2], Mahesh Kumar[2], Pradeep Kumar[1*]

[1]*School of Basic Sciences, Indian Institute of Technology Mandi, 175005, India*
[2]*Department of Electrical Engineering, Indian Institute of Technology Jodhpur, 342037, India*

**Abstract**

Excitonic quasi-particles, excitons/trions/bi-excitons, and their coupling with phonons and charge carriers play a crucial role in controlling the optical properties of atomically thin semiconducting 2D materials. In this work, we unravelled the dynamics of excitons/trions and their coupling with phonons and charge carriers in a few layers vertically and horizontally aligned MoS$_2$. We observed trion signature up to the highest recorded temperature (330 K) in both systems and have shown that the dynamics of excitons/trions and their coupling with phonons and electrons are more affected in vertically aligned MoS$_2$. A homogeneous linewidth broadening is observed with an increase in temperature. The linewidth broadening is attributed mainly to acoustic phonons in a low-temperature regime (<100 K). In contrast, acoustic and longitudinal optical phonons contributions to the linewidth broadening are observed at high temperature. We also observed the significant effects of interlayer coupling in both systems via understanding the temperature-dependent valence band splitting and trion binding energy. A decrease of 22 and 12% in valence band splitting with temperature rise is observed for the vertically and horizontally aligned MoS$_2$, respectively, suggesting that the valence band splitting is affected more in the case of vertically than horizontally aligned. Furthermore, we also notice a significant thermal quenching in the intensity of the trion band than that of exciton bands, attributed to the small binding energy of the trion.

[#]E-mail: deepu7727@gmail.com
[*]E-mail: pkumar@iitmandi.ac.in

# 1. Introduction

Two dimensional (2D) transition metal dichalcogenides (TMDCs) with atomic formula $MX_2$ (M= Mo, W and X=S, Se, Te) have emerged as promising materials for future electronic and optoelectronic device applications due to their tunable physical and optical properties [1]. $MoS_2$ is one of the crucial members of the layered TMDCs family, has attracted widespread attention because of potential applications for future electronic and optoelectronic devices. Monolayer $MoS_2$ shows a direct bandgap (1.8 eV) at the K point of the hexagonal Brillouin zone (BZ), while bulk $MoS_2$ possesses an indirect bandgap (1.2 eV) [2-3]. Monolayer $MoS_2$ show two strong photoluminescence (PL) emissions in the energy range of ~ 1.8-2.1 eV, attributed to the direct excitonic transition between the minima of the conduction band and the maxima of the split valence band at K point of BZ. Also monolayer or odd layers of $MoS_2$ show a strong spin-orbit coupling (SOC) due to inversion symmetry breaking [4-5]. This remarkable SOC coupled with crystal symmetry gives rise to the coupling between spin and valley degree of freedom, making the $MoS_2$ suitable for spintronic and valleytronic device applications. Additionally, this strong SOC lifts the spin degeneracy of the energy bands, resulting in the splitting of the valence band maxima and conduction band minima [6-7]. Further, at the K point of BZ, the splitting is prominent in the valence band (VB), while it is negligible in the conduction band (CB) for the $MoS_2$. The SOC and splitting results in many physical phenomena such as the spin Hall effect and valley Hall effect [8-9]. For the monolayer, only SOC plays a prominent role in the splitting of the VB. On the other hand, when the number of layers is tuned from the monolayer to few layers, the interlayer coupling between adjacent layer comes into the picture and play a finite role in VB splitting along with SOC. It was observed that the increase in thickness from monolayer to few layers increases VB splitting [10]. Besides, the 2D nature of the atomically thin $MoS_2$ leads to an increase in binding energy between electron and hole because of reduced dielectric screening effect, large effective mass, and strong quantum confinement, allowing the formation of a large variety of excitonic quasi-particles like excitons, charged excitons (trions) and bi-excitons [11-12].

In this work, we have carried out a comparative temperature-dependent PL study for the vertically (VA) and horizontally aligned (HA) few layers (~3 layers) of $MoS_2$ to understand the fundamental excitonic properties together with VB splitting, trion binding energy and thermal PL quenching as a function of temperature. Temperature-dependent PL spectroscopy has been widely used to understand the fundamental physics and dynamics of excitonic quasi-particle in TMDCs [13-17]. It is worth noting that most of the efforts have been devoted to study the HA $MoS_2$, while these studies for VA $MoS_2$ still lack so far. The relaxation of excitons or trions due to phonon interaction is vital for fundamental physics and potential device applications. Our study shows the significant role of acoustic phonons and longitudinal optical (LO) phonons in the relaxation of excitons/trions in VA and HA $MoS_2$.

## 2. Experimental Details

Few layer flakes of both VA and HA $MoS_2$ were obtained on a 300 nm $SiO_2$/Si substrate, using the chemical vapour deposition (CVD) technique [18]. Temperature-dependent PL measurements were carried out using the Horiba LabRAM HR evolution Raman spectrometer. An excitation wavelength of 532 nm was used to excite the PL spectrum. The laser power on the sample was kept very low, $\leq 1$ mW, to avoid any heating effect. A 50x long working distance objective lens was used to put incident light on the sample and collect the emitted light from the sample. The temperature variation was carried out using the closed cycle refrigerator (Montana) in a temperature range from 4 to 330 K, with a temperature accuracy of $\pm 0.1$ K. The waiting time for each PL measurement was ~10 min for better temperature stability.

## 3. Results and Discussions

Figures 1 (a) and 1 (b) show the temperature evaluation of the PL spectrum of both VA and HA $MoS_2$, respectively. The PL spectrum consists of a strong PL signal attributed to the A exciton, and a weak PL emission corresponds to the B exciton. Although the few layers of MoS$_2$ possesses an indirect bandgap, we observed a PL signal, unlike in the case of bulk MoS$_2$. The finite PL emission has also been seen for the others few layers of TMDCs [14,19]. The strong PL signals in a few layers of TMDC may appear due to direct bandgap hot luminescence [5]. As expected, redshifts, broadening and quenching in PL bands are observed as a function of increasing temperature in both VA and HA MoS$_2$. These effects could be clearly seen in the 2D colour contour maps of the PL intensity; see Figs 1 (c) and 1 (d) for the VA and HA, respectively.

To quantitatively understand the temperature-dependent energies, linewidths (FWHM) and intensities of the PL emission peaks, a sum of the Lorentzian function was used to extract the these parameters. The thick solid red lines in Figs. 1 (a) and 1 (b) show the total sum of Lorentzian fit, and the thin blue lines show individual peak fits. The excitonic feature A is associated with two unresolved energies bands $X_{A_-}$ and $X_{A_0}$. We assign the lower energy peak ($X_{A_-}$) to the negative trion ($A_-$), while the high energy peak ($X_{A_0}$) to the neutral exciton ($A_0$). At room temperature, peak position of $X_{A_-}$ and $X_{A_0}$ are located at ~ 1.81 and ~1.85 eV; ~ 1.81 and ~ 1.84 eV for VA and HA, respectively. The observed peak ($X_B$) corresponds to the B exciton, and the room temperature values are observed to be ~ 1.96 and 1.97 eV for VA and HA, respectively. The identification of trions and excitons signatures was done in line with the earlier reports on the MoS$_2$ [17, 20] and based on our estimated trion binding energy and VB splitting, discussed later. Excitons are the bound state of one electron and one hole and are the most thermally stable excitonic quasi-particles due to their large binding energy. When the excess number of electrons and holes are present, they can capture the photo-excited excitons to form charged excitons or trions. Trions are the three-particle bound states of either two electrons and one hole (negative trion) or one electron and two holes (positive trion).

Interestingly, we observed the trions signature up to the highest recorded temperature (330 K), indicating the stability of trions even up to room temperature, see Figs. 1 (a) and 1 (b). Figures 2 (a) and 2 (b) show the temperature dependence of energies and linewidths of the $X_{A_-}$, $X_{A_0}$ and $X_B$ bands for VA and HA MoS$_2$, respectively. With the rise in temperature, redshift and linewidth broadening are observed for all the bands. The decrease in energies of the bands as a function of increasing temperature may be attributed to the increase in the strength of exciton (trion)-electron-phonon interactions and thermal expansion of the lattice constant [15]. Generally, in semiconductors, the temperature-dependent energy of the bandgap is understood by two empirical relations. The first one was proposed by Varshni [21], and is given as $E(T) = E_0 - \frac{\alpha T^2}{(T+\beta)}$, where $E_0$ corresponds to the energy of the $X_{A_-}$, $X_{A_0}$ and $X_B$ bands at 0 K. $\alpha$ is the fitting parameter corresponding to the band energy temperature coefficient, which reflects the exciton (trion)-phonon interaction, the fitting parameter $\beta$ is associated with Debye temperature of the materials. The solid red lines in Figs. 2 (a) and 2 (b) illustrate the fitting curves to the $X_{A_-}$, $X_{A_0}$ and $X_B$ peaks using the Varshni relation, and a very good agreement to the experimental data is observed for both systems. The best-fit parameters extracted for the excitonic peaks $X_{A_-}$, $X_{A_0}$ and $X_B$ are listed in Table-I. For HA MoS$_2$, the values of $\beta$ are found to be ~ 495.6, ~ 631.1 and ~ 424.5 K for the bands $X_{A_-}$, $X_{A_0}$ and $X_B$, respectively; and ~ 751.7, ~ 524.8 and ~ 455.1 K are the estimated values of $\beta$ for the case of VA MoS$_2$. The obtained values of $\beta$ are much larger than those of estimated values for the A and B excitons in single-crystal of MoS$_2$ [22], while the average value of $\beta$ is comparable to that of monolayer and bulk MoS$_2$ obtained from first-principle calculations [23] and neutron scattering [24] measurements, respectively.

The second empirical relation to understanding the temperature-dependent bandgap was suggested by Donnell and Chen [25] and is given as $E(T) = E_0 - S*E_p [coth(\frac{E_p}{2k_bT}) - 1]$, where $k_b$ is the Boltzmann constant and $E_0$ corresponds to the energy of the excitonic bands at 0 K, $E_p = <\hbar\omega>$ is the average energy of the contributing phonons to the shift in energy bands with temperature. The factor S is a dimensionless Huang-Rhys factor describing the strength of exciton (trion)-electron-phonon coupling; a higher value of S reflects the stronger coupling. The solid green lines in Figs. 2 (a) and 2 (b) are the fitted curves to the experimental data using Donnell and Chen relation, and a good fitting is achieved to the experimental data. The best-fit parameters are listed in Table-II. The values of S are different for the $X_{A_-}$, $X_{A_0}$ and $X_B$ and may arise due to different coupling between excitons (trions) with phonons and electrons because of the different radius and effective mass of the excitons and trions [26-27]. The extracted value of the S factor is significant for $X_{A_-}$ and $X_B$ compared to that of $X_{A_0}$ in both HA and VA MoS$_2$, see Table-II. Further, we notice that the values of S corresponding to the excitonic bands are larger in VA than those of HA ones. The obtained values of $E_p$ are close to the energy reported for the backscattering forbidden optical phonon $E_{1g}$ (~ 35 meV), while these are almost half that of optical phonon $E_{2g}^1$ (~ 47 meV) and $A_{1g}$ (~ 50 meV) for MoS$_2$. Nevertheless, the obtained values of S and $E_p$ are in excellent agreement with previously reported values for the atomically thin MoS$_2$ [17, 28].

We now turn to the temperature-dependent linewidths (FWHM) of $X_{A_-}$, $X_{A_0}$ and $X_B$ bands. The linewidth of an exciton/trion could be strongly affected by the interactions of excitons/trions with LO phonons, acoustic phonons and free carriers. Moreover, presence of the impurities and imperfections also has a significant impact on the linewidth. Temperature dependence of the linewidths of $X_{A_-}$, $X_{A_0}$ and $X_B$ bands are shown in Figs. 2 (a) and 2 (b) for VA and HA MoS$_2$, respectively. A homogeneous linewidth broadening for all bands is observed with increasing temperature, which may be attributed to the coupling of the excitonic quasi-particles with phonons. At finite temperature, the linewidth of an excitonic band in semiconductors may be given as [29]

$$\Gamma(T) = \Gamma_0 + \Gamma_{AC}(T) + \Gamma_{LO}(T) = \Gamma_0 + \lambda_{AC}T + \lambda_{LO}N_{LO}(T) \qquad (1)$$

where the first term $\Gamma_0$ is the linewidth at 0 K, and it arises due to the scattering of excitons with impurities and imperfections. The second and third terms represent the contributions to the linewidth broadening due to the excitons scattering with acoustic and LO phonons, respectively. The coefficients $\lambda_{AC}$ and $\lambda_{LO}$ are the corresponding coupling strength of the excitons-acoustic phonons and excitons-LO phonons interactions, respectively. The interactions between excitons and LO phonons is understood by the Frohlich mechanism [30], while interactions with the acoustic phonons may be described by deformation-potential, which causes the lattice distortion [29]. $N_{LO}(T)$ is the Bose-Einstein distribution function representing population of the phonons at finite temperature, and is given as $N_{LO}(T) = 1/[e^{(E_{LO}/k_bT)} - 1]$; where $E_{LO}$ is the energy of the participating LO phonons. From the third term of Eq$^n$.1, apart from the estimation of Frohlich coupling strength, one can also estimate the average energy of the participating LO phonons, which play a predominant role in the linewidth broadening at high temperature. The solid blue lines are the fitted curves using the above Eq$^n$.1, and the fitting is in good agreement with the experimental data, see Figs. 2 (a) and 2 (b). The obtained best-fit parameters are listed in Table-III for both VA and HA. Figures 3 (a) and 3(b) describe the individual contributions from the acoustic and LO phonons to the linewidth of the excitonic bands for the VA and HA MoS$_2$, respectively. We observed that in the low-temperature regime (below 100 K), the linewidth broadening predominately results from the acoustic phonons, while the contributions from LO phonons is almost negligible. The minimal contributions to the linewidth from the LO phonons at low temperature may be understood as: energies of the LO phonons in MoS$_2$ lies in the range of ~ 30-50 meV, and in the low-temperature regime, energies of the LO phonons are more than that of thermal energy ($E_{LO} > k_b T$), giving in to a very small population of LO phonons, results in a negligible role of the LO phonons to the linewidth broadening. On the other hand, energies of the acoustic phonons are less than that of the thermal energy ($E_{AC} \leq k_b T$), making the acoustic phonons available in large to contribute to the linewidth broadening. Moreover, both acoustic and LO phonons role to the linewidth broadening is observed with increasing temperature above 100 K to the highest recorded temperature (330 K) in both VA and HA systems. In most of the semiconductors, including TMDCs, the prominent role of acoustic phonon to the linewidth broadening is either accounted only for in the low-temperature regime (< 100 K) or has been completely ignored because of the subtle role in linewidth at higher temperatures [31-34]. At the same time, a prominent role of acoustic phonons and LO phonons to the linewidth broadening is suggested up to or above the room temperature [35-36]. For the systems under study, i.e. VA and HA MoS$_2$, we observed that role of the acoustic phonon to the linewidth is larger in VA than HA, suggesting more lattice distortion in VA than HA with temperature variation. To further quantify the individual role of acoustic and LO phonons to the temperature-dependent linewidth broadening, we also adopted a fixed value of $E_{LO}$ ~ 47 meV, which corresponds to the energy of the $E_{2g}^1$ LO phonons in the MoS$_2$ [37], a similar procedure is also being followed in the literature [34]. The solid yellow lines in Figs. 2 (a) and 2(b) are the fitted curves using Eq$^n$.1 with fixed $E_{LO}$. Figures 3 (c) and 3 (d) show the individual contributions from acoustic and LO phonons to the linewidth of the excitonic peaks using the best fit values of the $\lambda_{AC}$ and $\lambda_{LO}$ obtained via fitting Eq$^n$.1 with fixed $E_{LO}$ for VA and HA, respectively. We note that it is in line with results obtained without fixing $E_{LO}$, see Figs. 3 (a) and 3 (b), suggesting the role of acoustic and LO phonons even at high temperature.

Figures 4 (a) and 4 (b) show the VB splitting and trion binding energy as a function of temperature for the VA and HA MoS$_2$, respectively. The VB splitting ($E_{X_B} - E_{X_{A_0}}$) at room temperature is ~ 120 and ~ 123 meV in the VA and HA MoS$_2$, respectively. The decrease in VB splitting is observed with increasing temperature in both VA and HA MoS$_2$. At the lowest recorded temperature (i.e. 4 K), the VB splitting is large in VA compared to that of the HA MoS$_2$, while the VB splitting is almost the same for both VA and HA MoS$_2$ at the highest recorded temperature. The observed decrease in VB splitting with temperature is ~ 22 and 12% for the VA and HA, respectively, suggesting that the VB splitting is affected more in VA than HA with temperature. The decrease in VB splitting with increasing temperature may be understood as: the increase in temperature reduces the interlayer coupling between adjacent layers due to increased interlayer spacing due to thermal expansion, resulting in a decrease in VB splitting. Temperature-dependent VB splitting may be well described using Bose-Einstein type expression [16, 38]. It may be given as $E_{X_B - X_{A_0}}(T) = E_{X_B - X_{A_0}}(0) - P/(e^{(Q/T)} - 1)$, where $E_{X_B - X_{A_0}}(0)$ is the VB splitting energy at 0 K, $P$ and $Q$ are the fitting parameters associated with the interlayer coupling strength and phonon temperature, respectively. The solid red lines are the fitting curves. The fitting is in very good agreement with the experimental data, see Figs. 4 (a) and 4 (b), and the best-fit parameters are found to be ($P$: 51 meV, $Q$: 353 K) and ($P$: 43 meV, $Q$: 448 K) for the VA and HA, respectively. We observed a large VB splitting in the VA than HA MoS$_2$ at low temperature, reflecting the large interlayer coupling in VA than HA at low temperature. The obtained value of $P$ is larger in VA than that of HA MoS$_2$ and is in good agreement with our above statement. Further, in the above assumption, we have considered that the interlayer coupling strength is independent of temperature, though it does depend on the temperature. As the temperature rise, the interlayer spacing between the layer will be increased due to thermal expansion, which leads to a decrease in interlayer coupling strength, resulting in a reduction of VB splitting, which is clearly seen in Figs. 4 (a) and 4(b) for both VA and HA, respectively.

We now focus on trion binding energy ($E_{A_0} - E_{A_-}$) and its dependence on the temperature. Temperature-dependent trion binding energy is shown in Figs. 4 (a) and 4(b) for VA and HA MoS$_2$, respectively. In our case, the energy difference between $X_{A_0}$ and $X_{A_-}$ peaks is found to be ~ 34-35 meV in both HA and VA at room temperature, which is in line with reported trion binding energy for thin layer MoS$_2$ [39-40], suggesting that the peak $X_{A_-}$ is originated due to trions contributions. Furthermore, we observed an increase in trion binding energy with rising temperature in both VA and HA, see Figs. 4 (a) and 4 (b). The interlayer coupling between the adjacent layers also plays a prominent role in the binding energy of the excitons/trions. The size of excitons/trions is strongly affected by tuning of interlayer spacing, and it increases with reducing interlayer spacing, resulting in a decrease in excitons/trions binding energy. An increase in trion binding energy with a temperature rise may be understood as: the increase in temperature leads to increased interlayer spacing because of thermal expansion, which may increase trion binding energy.

Further, we focus on the quenching in PL intensities of the excitonic bands with increasing temperature. Generally, the PL intensity of an excitonic band in semiconductors dramatically quenches with the rise in temperature due to thermally activated non-radiative recombination channels because of increased electron-phonon interactions. The thermal motion and population of the phonons increase with the rise in temperature, resulting in more scattering of photo-generated carriers via the non-radiative recombination process and reducing the probability of radiative transition, which gives rise to quenching in intensity with increasing temperature. For instance, quenching in exciton intensity occurs due to the decaying of excitons into electrons and holes, which may recombine with phonons via a non-radiative recombination process. The quenching in PL intensity with temperature rise could be clearly seen in the 2D colour contour plot as shown in Fig. 1 (c) and 1(d) for the VA and HA MoS$_2$, respectively. To quantitatively understand the temperature-dependent intensities of the PL bands, we have plotted the intensities of $X_{A_-}$, $X_{A_0}$ and $X_B$ bands as a function of temperature. Figures 4 (c) and 4(d) show the temperature dependence of the normalized PL intensities of $X_{A_-}$, $X_{A_0}$ and $X_B$ bands. We observe a significant intensity quenching for the trion peak ($X_{A_-}$) than the exciton peaks ($X_{A_0}$ or $X_B$) in both VA and HA with increased temperature. The significant quenching in intensity for $X_{A_-}$ than that of $X_{A_0}$ and $X_B$ with temperature rise, maybe arisen due to the smaller binding energy of trions than that of excitons. Furthermore, we also notice that the trion peak is strongly quenched in VA compared to HA one. The temperature-dependent PL intensity of a band in a semiconductor may be understood using the Arrhenius relation [41]. It may be given as $I(T) = I_0 / (1 + De^{(-E_a/k_bT)})$; where $I_0$ the intensity at 0 K, $E_a$ is the activation energy corresponding to the thermal quenching process and D is the constant fitting parameter. The solid red lines in Fig 4 (c) and Fig 4 (d) are the fitting curves, and the fitting is good in agreement with the experimental data. The thermal activation energy corresponding to the $X_{A_-}$ is found to be ~ 37.1 and ~ 21.3 meV for the VA and HA, respectively, indicating that trions in VA MoS$_2$ have more tendency to dissociate into neutral excitons than that of the HA, which we attribute to the electrons escaping from bound trions because of thermal fluctuations. The estimated $E_a$ for $X_{A_-}$ peak is found to be very close to values obtained for the average energy of the phonons, see Table-II. However, in our case, the estimation of $E_a$ for the $X_{A_0}$ and $X_B$ excitonic peaks using the Arrhenius relation is not possible because of the non-monotonically quenching in intensity with the rise in temperature. Nevertheless, an overall quenching in intensity is observed with the increase in temperature for the excitons $X_{A_0}$ and $X_B$ in both VA and HA MoS$_2$. Furthermore, a large quenching is observed for $X_{A_0}$ than that of $X_B$ in both systems. The decrease in intensity of excitons may be attributed to the dissociation of excitons into free carriers electrons and holes due to thermal fluctuations.

## 4. Conclusions

In summary, we carried out a detailed and comparative temperature-dependent photoluminescence study to understand the optical properties in a few layers vertically and horizontally aligned MoS$_2$ in a wide temperature range from 4 to 330 K. We observed the signatures of trions up to our highest recorded temperature with a room temperature binding energy of ~ 34-35 meV. Temperature-dependent thermal quenching, valence band splitting and trion binding energy, along with energy and linewidth of the excitonic quasi-particles, were understood here. Temperature-dependent valence band splitting and trion binding energy hint at the diagonally opposite role of interlayer interaction. Our finding suggests that the dynamics of excitonic quasi-particles are affected more in vertically than horizontally aligned MoS$_2$. We also extracted out the Debye temperature for both vertically and horizontally aligned MoS$_2$ via understanding the temperature-dependent energies of the excitonic bands.

**Acknowledgement:** PK acknowledges the Department of Science and Technology (DST) and IIT Mandi, India, for the financial support and IIT Mandi for providing the experimental facilities.

**Table-I** List of the fitting parameters for the VA and HA MoS$_2$ obtained using Varshni expressions described in the text.

| Peak | VA-MoS$_2$ | | | HA-MoS$_2$ | | |
|---|---|---|---|---|---|---|
| | $E_0$ (meV) | $\alpha$ $10^{-4}$ (eV K$^{-1}$) | $\beta$ (K) | $E_0$ (meV) | $\alpha$ $10^{-4}$ (eV K$^{-1}$) | $\beta$ (K) |
| X$_{A-}$ | 1882.1±0.02 | 8.1±0.6 | 751.8±77.3 | 1859.5±0.05 | 5.3±0.1 | 495.6±40.3 |
| X$_{A0}$ | 1911.5±0.02 | 5.7±0.3 | 524.8±41.4 | 1886.8±0.05 | 5.9±0.2 | 631.1±50.9 |
| X$_B$ | 2057.3±0.03 | 7.1±0.3 | 455.1±32.6 | 2024.9±0.02 | 5.2±0.2 | 424.5±29.0 |

**Table-II:** List of the fitting parameters for the VA and HA MoS$_2$ obtained using Donnell and Chen's expressions described in the text.

| Peak | VA-MoS$_2$ | | | HA-MoS$_2$ | | |
|---|---|---|---|---|---|---|
| | $E_0$ (meV) | S | $E_p$ (meV) | $E_0$ (meV) | S | $E_p$ (meV) |
| X$_{A-}$ | 1879.8±0.07 | 2.26±0.13 | 27.2±2.2 | 1856.5±0.08 | 1.79±0.15 | 24.5±2.9 |
| X$_{A0}$ | 1909.5±0.08 | 1.94±0.15 | 24.8±2.8 | 1886.3±0.06 | 1.51±0.09 | 18.8±2.1 |
| X$_B$ | 2054.5±0.09 | 2.60±0.08 | 23.7±1.4 | 2022.9±0.09 | 1.93±0.12 | 23.2±3.4 |

**Table-III:** Values of the parameters, extracted by fitting eqn.1, describing the temperature dependence of the broadening of the observed excitonic peaks in VA and HA MoS$_2$.

| Peak | VA-MoS$_2$ | | | | HA-MoS$_2$ | | | |
|---|---|---|---|---|---|---|---|---|
| | $\Gamma_0$ (meV) | $\lambda_{AC}$ (µeVK$^{-1}$) | $\lambda_{Lo}$ (meV) | $E_{Lo}$ (meV) | $\Gamma_0$ (meV) | $\lambda_{AC}$ (µeVK$^{-1}$) | $\lambda_{Lo}$ (meV) | $E_{Lo}$ (meV) |
| X$_{A-}$ | 52.2±0.5 | 51±4 | 56.3±5 | 90.4±6.0 | 41.3±0.4 | 56±5 | 113.8±13 | 81.1±4.6 |
| X$_{A0}$ | 46.4±0.2 | 95±3 | 171.6±8 | 66.8±3.8 | 44.9±1.5 | 88±2 | 333.0±9 | 64.7±7.2 |
| X$_B$ | 92.1±0.4 | 175±5 | 293.9±10 | 57.5±5.6 | 102.7±1.0 | 94±1 | 63.4±5 | 63.8±4.8 |

**FIGURES:**

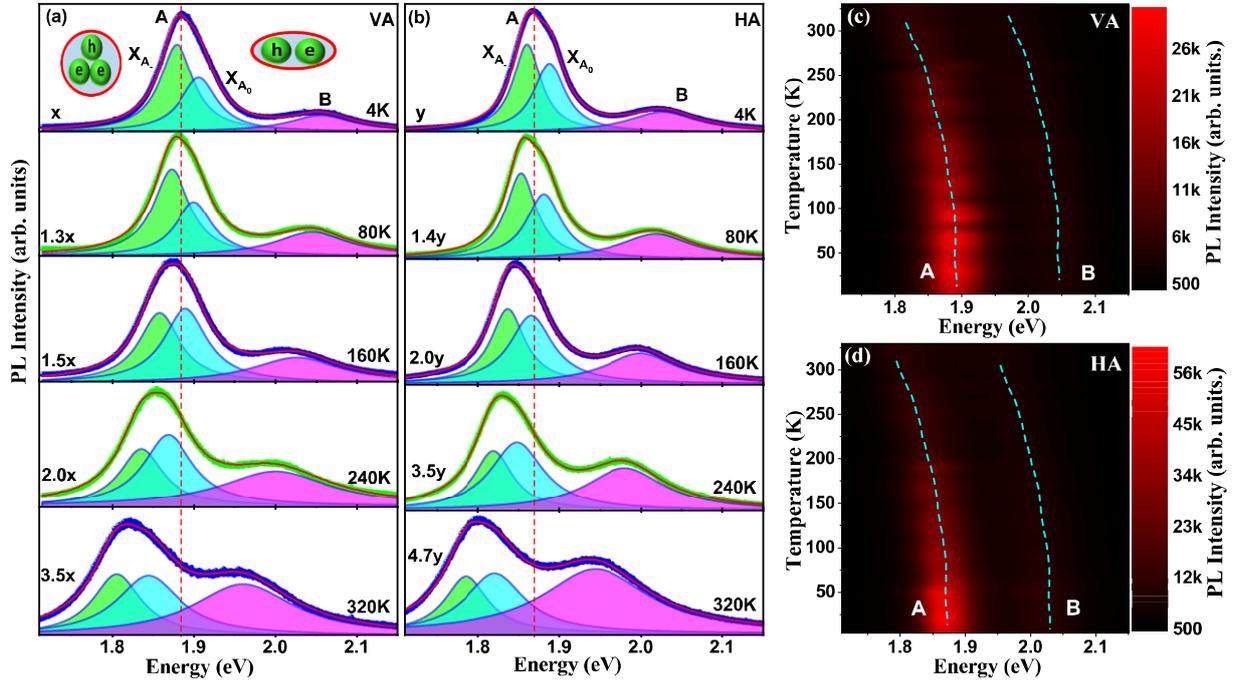

**FIGURE 1:** (a) and (b) Temperature evaluation of the PL spectra for VA and HA MoS$_2$, respectively. Insets show the schematic representation of the negative trion and exciton quasi-particle bound states. (c) and (d) 2D colour contour maps of the PL intensity versus PL energy as a function of temperature for VA and HA MoS$_2$, respectively. The dotted lines show the shift in energies of the PL peaks with temperature.

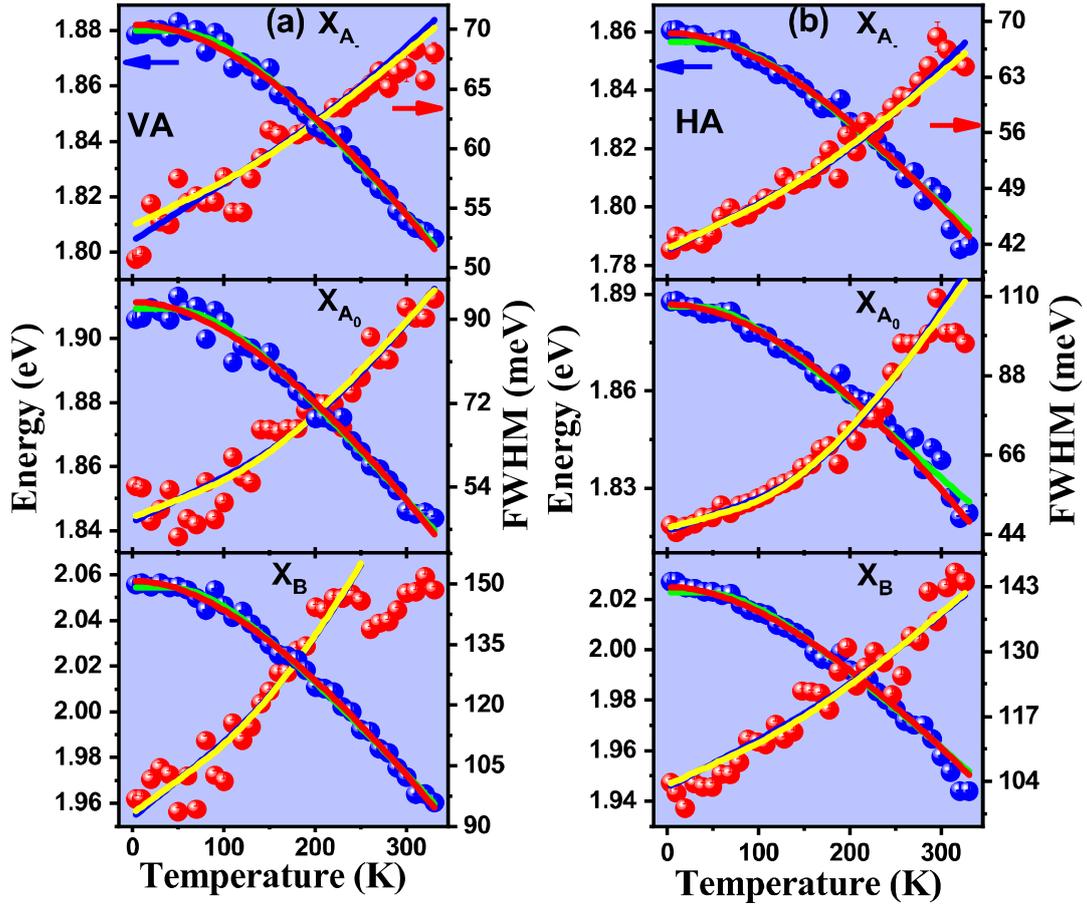

**FIGURE 2:** (a) and (b) Temperature dependence of the energies and FWHM of the $X_{A_-}$, $X_{A_0}$ and $X_B$ peaks for the VA and HA MoS$_2$, respectively. The solid lines are the fitted curves as described in the text.

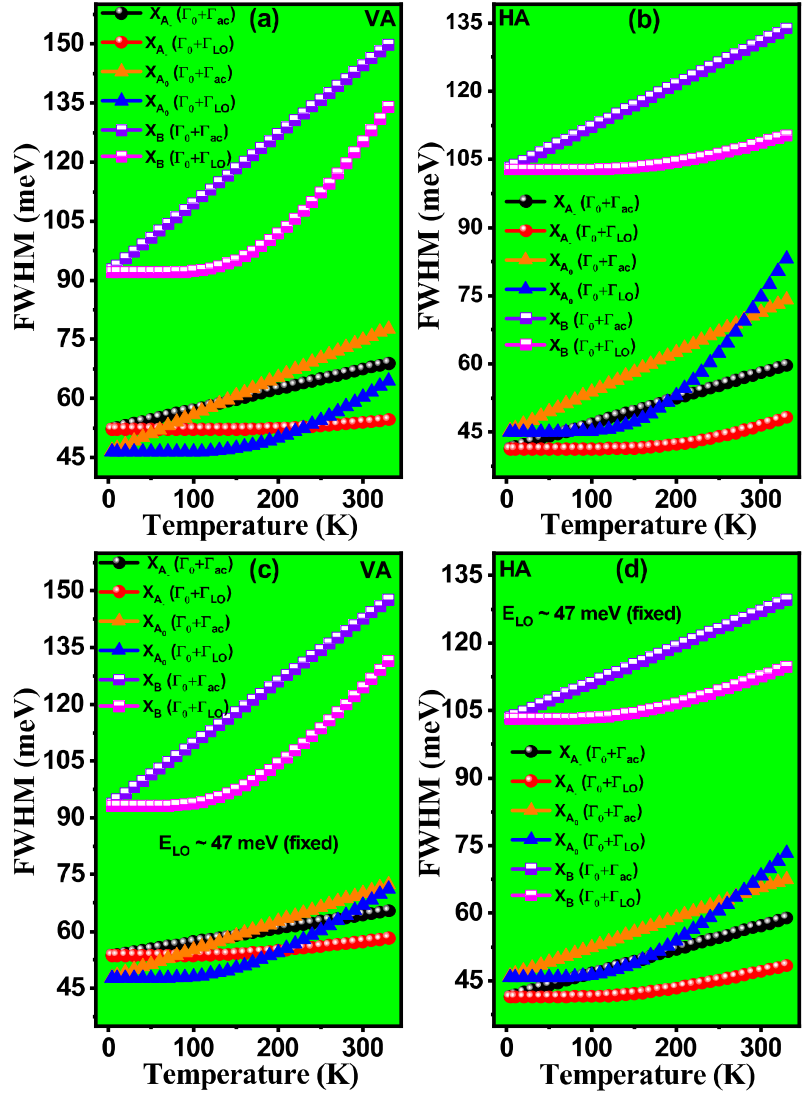

**FIGURE 3:** (a) and (b) Individual contributions to the linewidth of the $X_{A_-}$, $X_{A_0}$ and $X_B$ peaks from acoustic and LO phonons for the VA and HA MoS$_2$, respectively. (c) and (d) show the individual contributions from acoustic and LO phonons to the linewidth of the excitonic peaks using the best fit values of the $\lambda_{AC}$ and $\lambda_{LO}$ obtained via fitting Eq$^n$.1 with fixed $E_{LO}$ for VA and HA MoS$_2$, respectively.

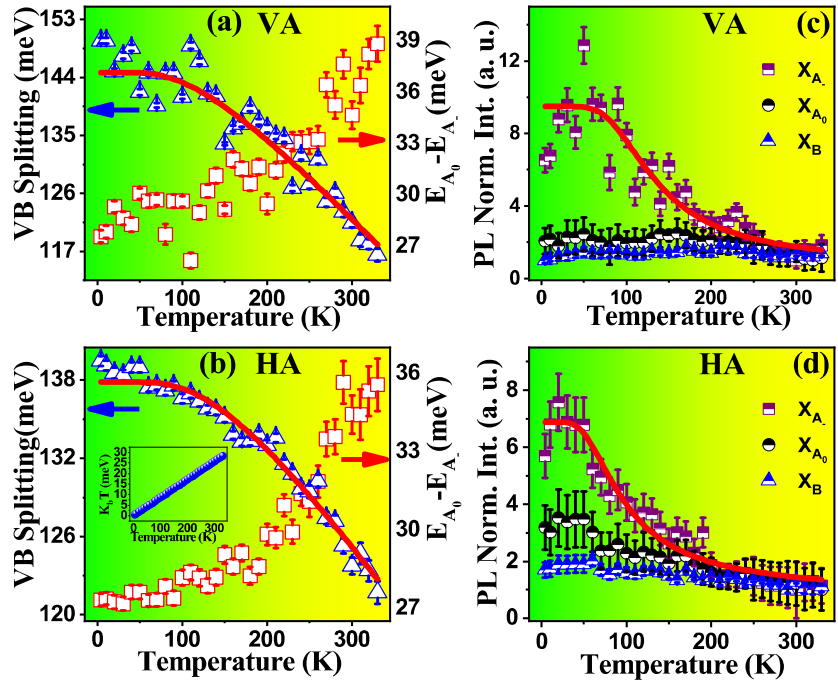

**FIGURE 4:** (a) and (b) Temperature dependence of the VB splitting (blue triangle) and trion binding energy (red square) for the VA and HA MoS$_2$, respectively. The inset in (b) is the evaluation of thermal energy with temperature. (c) and (d) Temperature-dependent normalized PL intensities corresponds to the $X_{A_-}$, $X_{A_0}$ and $X_B$ excitonic peaks in the VA and HA MoS$_2$, respectively.